\documentclass[12pt]{iopart}
\usepackage{iopams}
\usepackage{setstack}
\usepackage{graphicx}
\usepackage{cite}
\usepackage{epsfig}

\begin{document}

\centerline{\title{End of a Dark Age?}}

\author{W.M. Stuckey$^1$, Timothy McDevitt$^2$, A.K. Sten$^3$ and Michael Silberstein$^4$}

\address{$^1$ Department of Physics \\ Elizabethtown College \\ Elizabethtown, PA  17022 \\stuckeym@etown.edu}
\address{$^2$ Department of Mathematical Sciences \\ Elizabethtown College \\ Elizabethtown, PA  17022 \\ mcdevittt@etown.edu}
\address{$^3$ Department of Physics \\ Elizabethtown College \\ Elizabethtown, PA  17022 \\stena@etown.edu}
\address{$^4$ Department of Philosophy and Foundations of Physics \\ Committee for Philosophy and the Sciences \\ University of Maryland \\ College Park, MD  20742 \\ msilbers@umd.edu}

\begin{abstract}
We argue that dark matter and dark energy phenomena associated with galactic rotation curves, X-ray cluster mass profiles, and type Ia supernova data can be accounted for via small corrections to idealized general relativistic spacetime geometries due to disordered locality. Accordingly, we fit THINGS rotation curve data rivaling modified Newtonian dynamics, ROSAT/ASCA X-ray cluster mass profile data rivaling metric-skew-tensor gravity, and SCP Union2.1 SN Ia data rivaling $\Lambda$CDM without non-baryonic dark matter or a cosmological constant. In the case of dark matter, we geometrically modify proper mass interior to the Schwarzschild solution. In the case of dark energy, we modify proper distance in Einstein-deSitter cosmology. Therefore, the phenomena of dark matter and dark energy may be chimeras created by an errant belief that spacetime is a differentiable manifold rather than a disordered graph.
\end{abstract}
\thispagestyle{empty}

\clearpage

\section{Introduction}

Gravitational theory entered its current ``dark age'' in the early 1930’s when galactic rotation curves (RC's) and galactic cluster masses were observed to deviate from Newtonian expectations based on luminous matter and mass-to-luminosity ratios \cite{oort,zwicky1,zwicky2,rubin}. These are two aspects of the dark matter phenomenon \cite{garrett}, a phenomenon that was followed much more recently by ``the discovery of the accelerating expansion of the Universe through observations of distant supernovae \cite{nobel},'' ushering in the phenomenon of dark energy. Both non-baryonic dark matter (DM) and a cosmological constant $\Lambda$ (dark energy) play important roles in the concordance model $\Lambda$CDM \cite{planck} where baryonic matter comprises $\sim$4\% of all the energy density in the universe, DM comprises $\sim$23\%, and dark energy comprises $\sim$73\%. The sum of these contributions results in a spatially flat, radiation-dominated universe transitioning to a spatially flat, matter-dominated universe which does a good job accounting for cosmological observations, e.g., anisotropies in the power spectrum of the CMB \cite{hu} and galactic distributions attributed to baryon acoustic oscillations \cite{eisenstein}. However, 80 years after first being posited, there is still no independent verification of DM and galactic RC's do not conform to the theoretical predictions of $\Lambda$CDM for the distribution of DM on galactic scales \cite{gentile1}. After listing DM's attributes, i.e., dark, cold, abundant, stable, and dissipationless, Sean Carroll concludes \cite{carroll1}, ``So should we be surprised that we live in a universe full of dark matter? I’m going to say: yes.'' Likewise, we have no consensus explanation for $\Lambda$ of the size needed in $\Lambda$CDM \cite{carroll2,weinberg1,bianchi}. In other words, while DM and $\Lambda$ serve us well in $\Lambda$CDM, they are also problematic. ``As Tom Shanks once said, there are only two things wrong with $\Lambda$CDM: $\Lambda$ and CDM'' \cite{bull}. For these reasons and others, there are efforts to explain gravitational phenomena on astrophysical scales without DM or $\Lambda$ \cite{garfinkle,paranjape,tanimoto,clarkson,milgrom1,milgrom2,sanders1,bekenstein,sanders2,zlosnik,zhao,blanchet,brownstein1,brownstein2}. However, as far as we know, there is no attempt to get rid of both DM and $\Lambda$, which is what we propose.
 
Concerning our motivation for explaining dark matter and dark energy phenomena, we point out that we came to these problems from the foundations of physics. Our proposed fundamental ontology, Relational Blockworld (RBW), was originally conceived as an interpretation of quantum mechanics \cite{stuckey1,stuckey2,silber1}, but it quickly became apparent that it has implications for quantum gravity, unification and astrophysics \cite{silber2,stuckey3}. According to RBW, reality is fundamentally discrete, so although the lattice geometry of Regge calculus \cite{regge,misner,barrett,williams} is typically viewed as an approximation to the continuous spacetime manifold of general relativity (GR), it could be that discrete spacetime is fundamental while ``the usual continuum theory is very likely only an approximation'' \cite{feinberg} and that is what we assume. Further, the links of a Regge calculus graph can connect non-neighboring points of the corresponding GR spacetime manifold leading to small corrections to the corresponding GR spacetime geometry. The direct connection between non-neighboring points on the spacetime manifold is referred to as ``disordered locality'' \cite{caravelli} and has been used on astrophysical scales to explain dark energy \cite{prescod}. Our views deviate from the standard use of Regge calculus, so we refer to our approach as modified Regge calculus (MORC). Thus, ours is a foundationally motivated approach to the problems of dark matter and dark energy.

\section{The Model}

RBW's disordered locality is a variation on the old idea of direct particle interaction \cite{wheeler,hawking,davies1,davies2,hoyle,narlikar} whereby the na$\ddot{\mbox{\i}}$ve notion of a mediating quantum field between sources is eliminated. A discrete graphical spacetime is obviously going to create conflict with the differentiable spacetime manifold of GR, but particularly so when quantum matter-energy exchange occurs between sources at distances exceeding the validity of a flat spacetime approximation. Thus, we propose that Regge calculus be modified by adding links between non-neighboring points in the context of the corresponding continuous spacetime manifold. One would then solve Regge's equations for the lattice modified per disordered locality. Of course, without some highly symmetric form of disordered locality, we expect the modified Regge's equations would have to be solved numerically. Thus, we assume that the existence of modest disordered locality in the exact Regge calculus graph justifies small corrections to the corresponding approximate GR solution. In practice we imagine modified Regge calculus graphs with greatly simplifying assumptions would be used as approximations of the exact Regge calculus graph (also true of standard Regge calculus, obviously). When all link lengths are small, i.e., in the absence of disordered locality, these approximate Regge calculus solutions would then correspond to GR solutions and we have standard Regge calculus \cite{brewin1,miller,brewin2}. As with GR solutions, Regge calculus solutions are nontrivial and there is no reason to believe that finding extrema of a Regge graphical action modified per disordered locality would be any easier. Rather, at this point, we are simply operating on the assumption that a modified Regge graphical action and its extrema will make correspondence with Regge calculus and GR in the proper limits. Motivated by RBW's prediction of disordered locality, we are systematically exploring possible geometric corrections to astrophysical phenomena that may be examples of disordered locality. It seems to us that dark energy and dark matter are two such examples. If we can find simple geometric corrections that resolve the problems of dark energy and dark matter, then we will use these as guides to produce a simplified cosmological Regge graphical action modified per disordered locality.
 
Accordingly, we introduced simple geometric corrections to idealized GR spacetime structure on large scales to account for observational data associated with dark matter, i.e., galactic RC's and galactic cluster mass profiles, and dark energy, i.e., type Ia supernova data. First, we fit the SCP Union2.1 supernova data matching that of $\Lambda$CDM via a simple correction of proper distance in Einstein-deSitter (EdS) cosmology \cite{stuckey4,stuckey5}. Specifically,
 
\begin{equation}
D_{L}=(1+z)D_{p} \rightarrow (1+z)D_{p} \sqrt{1+\frac{D_{p}}{A}}
\end{equation}
where $D_L$ is the luminosity distance, $z$ is the redshift, $D_p$ is the proper distance obtained using the Regge calculus EdS solution, and $A$ is a fitting parameter. From our Regge calculus EdS solution we have
\begin{equation}
D_p = \int \left(\frac{F'(b)}{bF(b)}\sqrt{1+\frac{b^2}{4}}\right)db
\end{equation}
where
\begin{equation}
F(b) = \frac{\sqrt{4+b^2}}{2[\pi -cos^{-1}(\frac{b^2}{4+2b^2})-2cos^{-1}(\frac{\sqrt{4+3b^2}}{2\sqrt{2+b^2}})]}
\end{equation}
with $b = \frac{R\dot{a}}{c}$. 
The type Ia supernova data to be fit is distance modulus ($\mu$) versus redshift ($z$), i.e., $\mu = 5 \log \left(\frac{D_L}{10pc}\right)$. The MORC sum of squares error (SSE) for $\frac{\mu}{5} - 8$ is robust against variation in coordinate lattice spacing $R$ and nodal mass $m$. We find the MORC best fit SSE = (1.630 $\pm$ 0.002) for $A$ = (7.48 $\leftrightarrow$ 10.6) Gcy with a current Hubble constant of $H_o$ = (69.9 $\leftrightarrow$ 75.1) km/s/Mpc using the MORC values $R$ = (2.11 $\leftrightarrow$ 8.39) Gcy and $m$ = (0.301 $\leftrightarrow$ 17.5) x 10$^{51}$ kg. The best fit $\Lambda$CDM gave SSE = (1.639 $\pm$ 0.003) using $H_o$ = (68.9 $\leftrightarrow$ 70.1) km/s/Mpc, $\Omega_M$ = (0.24 $\leftrightarrow$ 0.28) and $\Omega_\Lambda$ = (0.72 $\leftrightarrow$ 0.76). Both of these fits were superior to the EdS best fit with SSE = 2.67 and $H_o$ = 60.9 km/s/Mpc (Figure \ref{fig1}). A recent study has found $H_o$ = 73.00 $\pm$ 1.75 km/s/Mpc \cite{riess}. [For details see references [51] and [52]. Note: Those fits were for the older SCP Union2 SN Ia data.]
 
In order to account for dark matter phenomena, we note that in addition to a graphical spacetime with disordered locality, RBW assumes relationalism/contextuality \cite{auffeves}, i.e., mass is not an intrinsic property of matter, but is rather a characterization of spacetime geometry, itself a system of relations. As such, matter can simultaneously have different values of mass, each different value of mass associated with a different spacetime context. Like disordered locality, contextuality is not new to physics, e.g., it already exists in GR. Specifically, it is well known that the mass of the matter interior to the Schwarzschild solution (proper mass) can differ from the dynamic mass of that same matter per the exterior Schwarzschild metric \cite{wald,stuckey6}. While it may seem unnecessary to bring GR to bear on such rarified distributions of matter with non-relativistic rotation speeds ($v \ll c$), Cooperstock {\it et al.} used GR instead of Newtonian gravity in fitting galactic RC's and found that the non-luminous matter in galaxies ``is considerably more modest in extent than the DM extent claimed on the basis of Newtonian gravitational dynamics'' \cite{magalhaes,carrick,cooperstock}. Thus, we used contextuality and disordered locality to motivate a simple geometric modification to proper mass to obtain dynamic mass for MORC fits \cite{stuckey7} of twelve high-resolution galactic RC's from The HI Nearby Galaxy Survey \cite{walter} (THINGS) used by Gentile {\it et al}. to explore modified Newtonian dynamics (MOND) fits \cite{gentile2}. Specifically, we modified the proper mass $\Delta M_p$ of each (discrete) annulus of galactic matter to obtain its dynamic mass $\Delta M$ ($i^{th}$ component, where bulge, disk, and gas are the possible components) per

\begin{equation}
\Delta M_i=\delta_i (\frac{r_2+r_1}{2})^\xi \Delta M_p \label{RCFit}
\end{equation}
with $\delta_i$ and $\xi$ (same for all components) fitting parameters. The THINGS data to be fit is rotation velocity versus orbital radius. Gentile {\it et al}. describe these data as ``the most reliable for mass modelling, and they are the highest quality RC's currently available for a sample of galaxies spanning a wide range of luminosities.'' MORC fits rival MOND fits which were deemed ``very successful'' for these data \cite{gentile2} (Figure \ref{fig2}).

Finally, we used this same technique to fit the mass profiles of the eleven X-ray clusters found in Brownstein \cite{brownstein3} as obtained from Reiprich and B$\ddot{\mbox{o}}$hringer \cite{reiprich1,reiprich2} using combined ROSAT (ROentgen SATellite) and ASCA (Advanced Satellite for Cosmology and Astrophysics) data. Specifically, we used the continuum version of Eq. (\ref{RCFit})
 
\begin{equation}
M(r)=\int_{0}^{r}\delta r^\xi dM_p=4\pi \int_{0}^{r}\delta r'^{\xi } \rho (r')r'^2dr'
\end{equation} 
to modify the proper mass of each annulus of intracluster medium gas to obtain its dynamic mass. The X-ray cluster mass profile data to be fit is mass versus radius. MORC fits rival metric-skew-tensor gravity (MSTG) fits which bested MOND and scalar-tensor-vector gravity fits of these same data \cite{brownstein3} (Figure \ref{fig3}). [For details on dark matter fits see reference \cite{stuckey7}.]

\section{Conclusion}

The fundamental ontological entity per RBW is the ``spacetimesource element.'' A spacetimesource element is $of$ space and time, not $in$ space and time, and its properties are determined relationally and contextually per its classical context. The distribution of spacetimesource elements in their classical context is given by an adynamical global constraint that underwrites quantum physics. This view leads to a modified Regge calculus (MORC). We have shown that dark matter and dark energy phenomena associated with galactic rotation curves, X-ray cluster mass profiles, and type Ia supernova data can be accounted for via small corrections to GR spacetime geometries motivated by RBW’s disordered locality and relationalism/contextuality as implemented by MORC. Accordingly, the phenomena of dark matter and dark energy may be chimeras created by an errant belief that spacetime is a differentiable manifold instead of a disordered graph.

\section*{References}

\begin{figure}[h]
\begin{center}
\includegraphics[height=60mm]{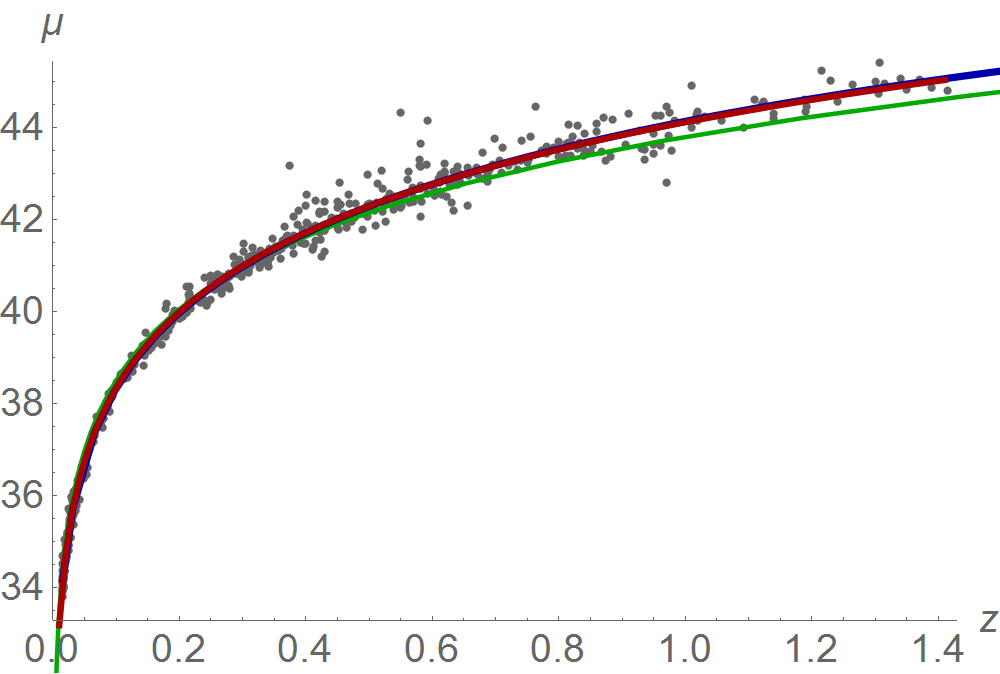} \label{fig1}
\end{center}
\caption{\textbf{Type Ia Supernova Data:} Plot of SCP Union2.1 SN Ia data (distance modulus versus redshift) along with the best fits for EdS (green), $\Lambda$CDM (blue), and MORC (red). The MORC curve is terminated at $z$ = 1.4 in this figure so that the $\Lambda$CDM curve is visible underneath.}
\label{fig1}
\end{figure}

\begin{figure}[h]
\begin{center}
\includegraphics[height=60mm]{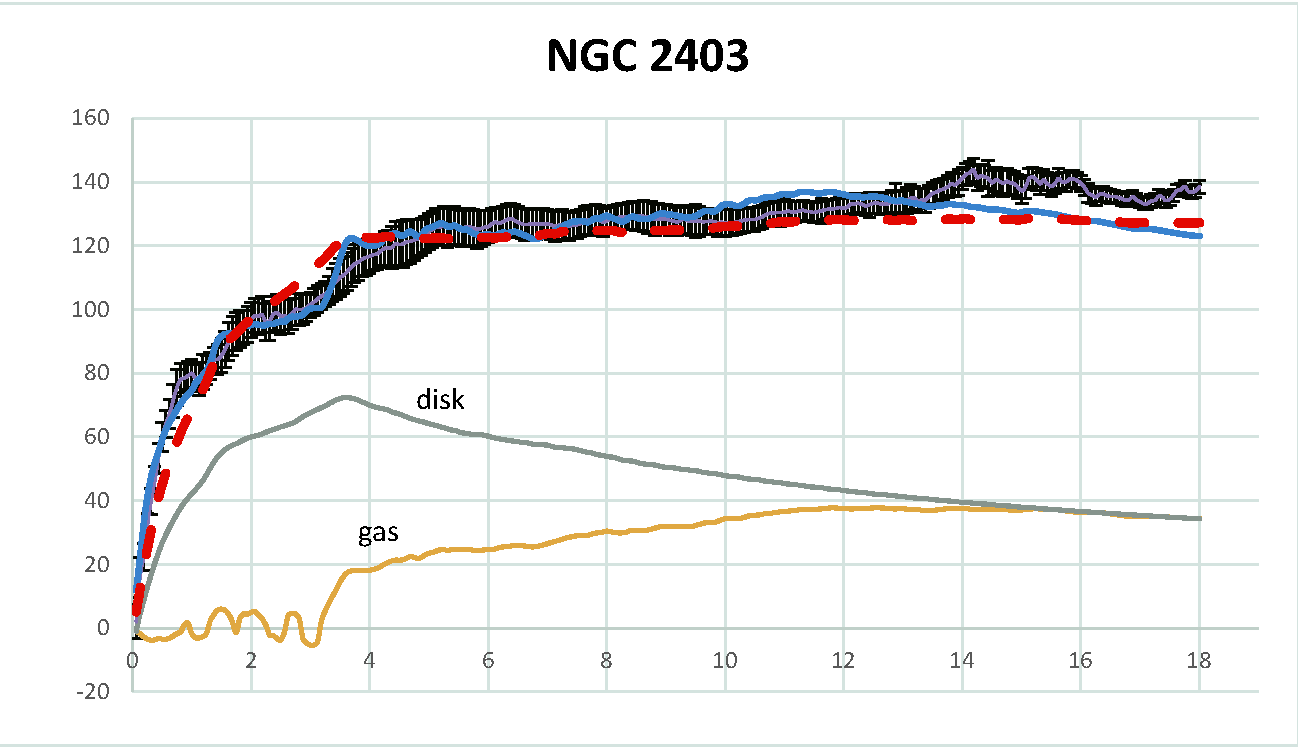} \label{fig2}
\end{center}
\caption{\textbf{Galactic Rotation Curves:} This and the following eleven graphs show MORC (solid blue) and MOND (dashed red) fits of THINGS galactic RC's (with error bars). Disk, gas and bulge curves are labeled. Bulge curves are not always available. Vertical axis is rotation velocity in km/s and horizontal axis is orbital radius in kpc. The mean square error (MSE) for this fit is 12.1 (km/s)$^2$ for MORC and 34.4 (km/s)$^2$ for MOND.
\linebreak
\linebreak
Other eleven graphs at http://arxiv.org/abs/1509.09288}
\label{fig2}
\end{figure}

\begin{figure}[h]
\begin{center}
\includegraphics[height=60mm]{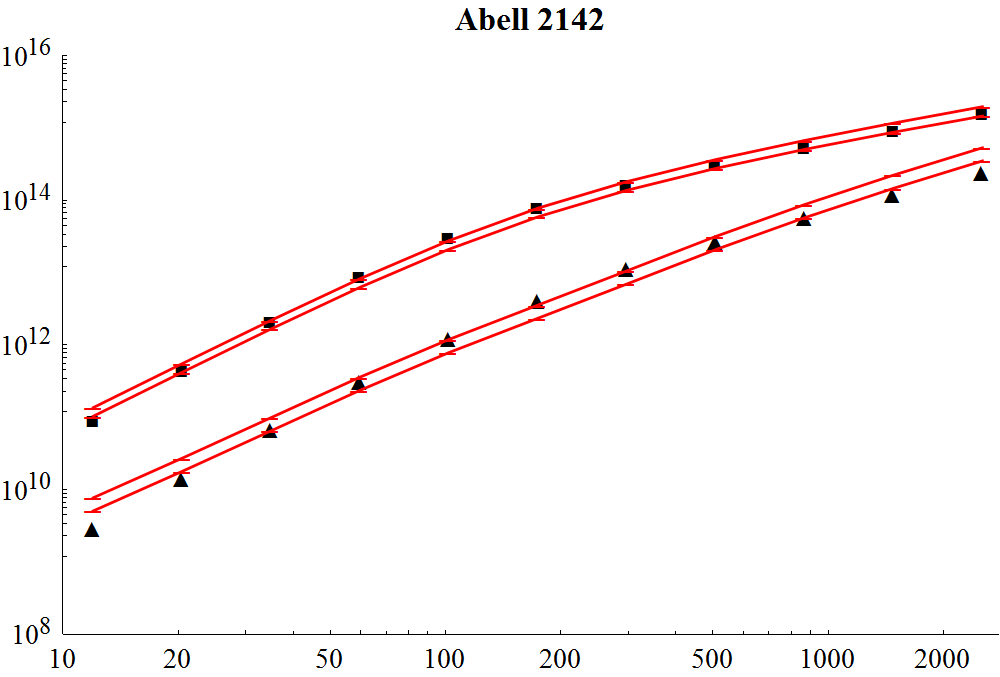} \label{fig3}
\end{center}
\caption{\textbf{Mass Profiles of X-ray Clusters:} This and the following ten log-log plots show MORC and MSTG fits of X-ray cluster mass profiles (compiled from ROSAT and ASCA data). Vertical scale is in solar masses and horizontal scale is in kpc. MORC is increasing the gas mass (triangles) to fit the dynamic mass (squares). MSTG is decreasing the dynamic mass to fit the gas mass. The sizes of the objects are approximately equal to their errors. MORC fit is the upper pair of lines (connecting fit points) over the squares where line separation corresponds to error. MSTG fit is the lower pair of lines (connecting fit points) over the triangles where line separation corresponds to error. MSE is ($\Delta$ $\log{(M)})^2$ image and for this fit MORC MSE = 0.00369 and MSTG MSE = 0.0302.
\linebreak
\linebreak
Other ten graphs at http://arxiv.org/abs/1509.09288}
\label{fig3}
\end{figure}
\end{document}